\documentclass[12pt]{article}
\usepackage{geometry}
\usepackage{xcolor}
\usepackage{a4}
\usepackage{graphicx}
\usepackage{epsf}
\usepackage{amsmath}
\usepackage{amssymb}
\usepackage{cite}
\newcommand{\be}{\begin{equation}}
\newcommand{\ee}{\end{equation}}

\newcommand{\Rmnum}[1]{\expandafter\@slowromancap\romannumeral #1@}
\newcommand{\bea}{\begin{eqnarray}}
\newcommand{\eea}{\end{eqnarray}}
\begin{document}
\def\C{{\mathbb{C}}}
\def\R{{\mathbb{R}}}
\def\s{{\mathbb{S}}}
\def\T{{\mathbb{T}}}
\def\Z{{\mathbb{Z}}}
\def\W{{\mathbb{W}}}
\def\Bbb{\mathbb}
\def\BZ{\Bbb Z} \def\BR{\Bbb R}
\def\BW{\Bbb W}
\def\BM{\Bbb M}
\def\BC{\Bbb C} \def\BP{\Bbb P}
\def\CP{\BC\BP}
\begin{titlepage}
\title{Galactic Dark Matter and Bertrand Space-times} \author{} 
\date{
Dipanjan Dey, Kaushik Bhattacharya, Tapobrata Sarkar 
\thanks{\noindent 
E-mail:~ deydip, kaushikb, tapo @iitk.ac.in} 
\vskip0.4cm 
{\sl Department of Physics, \\ 
Indian Institute of Technology,\\ 
Kanpur 208016, \\ 
India}} 
\maketitle 
 
\abstract{Bertrand space-times (BSTs) are static, spherically symmetric solutions of Einstein's equations, that admit stable, closed
orbits.  Starting from the fact that to a good approximation, stars in the disc or halo regions of typical galaxies move in such orbits, we propose that, 
under certain physical assumptions, the dark matter distribution of some low surface brightness (LSB) galaxies can seed a particular class of BSTs. In the
Newtonian limit, it is shown that for flat rotation curves, our proposal leads to an analytic prediction of the Navarro-Frenk-White (NFW) 
dark matter profile \cite{nfwprof}. We further show that the dark matter distribution that seeds the BST, is described by a two-fluid anisotropic model, and 
present its analytic solution. A new solution of the Einstein's equations, with an internal BST and an external Schwarzschild metric, is also constructed.}
\end{titlepage}

\section{Introduction}
\label{intro}

A Bertrand space-time \cite{perlick} is defined as one in which each point in a spatial hypersurface admits closed, stable
orbits. This generalizes the well known Bertrand's theorem in classical mechanics \cite{bert},\cite{goldstein} to a general
relativistic scenario (for related work in the special relativistic case, see \cite{kb}). BSTs
are static, spherically symmetric solutions of Einstein equations, which require a finite energy-momentum tensor, unlike
the Schwarzschild solution. Till now, most of the work on BSTs
\cite{enciso} have studied the geometrical nature of these space-times, e.g. the nature of classical and quantum operators
associated with these. The astrophysical nature of these space-times have been less explored. One particular reason for this may be due to the fact that
BSTs can break certain energy conditions in general relativity as noted in \cite{perlick}. But in a previous work \cite{kbs}, it
was shown that one class of BSTs (which we call BSTs of type II), does respect the various energy conditions of GR
if the parameters of the space-time are appropriately chosen.

The mathematical properties of BSTs literally beg for a physical interpretation. Indeed, it is not difficult to imagine that an
appropriate context in which BSTs would be important might be astrophysical in nature, as, up to a good approximation, stars move in
Keplerian orbits in certain galactic regions \cite{binneytremaine}. This is the main idea that we develop in this paper and our main result is
that Bertrand space-times may be seeded by galactic dark matter. Specifically, we show that the assumption that stars and other luminous 
matter follow time-like geodesics in BSTs, leads to analytic derivations of some well known galactic dark matter density profiles. We further show 
that the matter which seeds our BST is composed of two perfect fluids, and their exotic nature, in addition to the form of their density profiles, will 
lead us to the interpretation that the matter seeding the BSTs should be dark matter. 

This paper is organized as follows. In section 2, we present some generalities of BSTs to set the notations and conventions used
in the rest of the paper. In particular, we compute the circular velocity for time-like geodesics, show that these can be used to 
model galactic rotation curves, and compare with data from some LSB galaxies. We also use this to calculate the Newtonian density profile, and
show that the NFW and Hernquist profiles arise as natural consequences from the form of variation of circular velocity with radial distance. 
In section 3, we first present the 
energy momentum tensor for BSTs, analyze the weak energy condition, and motivate the need for a two fluid model to describe the matter distribution 
associated with our BST. The two-fluid model is solved in the subsequent subsection. In section 4, apart from some general comments and discussions, 
we indicate some preliminary results on a different possible form of BST. We also present a new 
space-time that is internally a BST and externally Schwarzschild. Section 5 ends this work with conclusions and possible directions for future research.
 
\section{Bertrand Space-times: General Considerations}

Formally, the definition of a Bertrand space-time \cite{perlick}, \cite{enciso} arises via a static, spherically symmetric Lorentzian
manifold $(M, g)$ whose domain is diffeomorphic to a product manifold $(r_1\,,\,r_2)\times S^2\times \mathbb{R}$ with the metric $g$ given
by ($c$ is the speed of light) 
\begin{eqnarray}
ds^2 = -e^{2\nu(r)}c^2 dt^2 + e^{2\lambda(r)} dr^2 + r^2 (d \theta^2 +
\sin^2 \theta d \phi^2)\,,
\label{invl}
\end{eqnarray}
where $r$ ranges in the open interval $(r_1\,,\,r_2)$ with $0\leq r_1 < r_2 \leq \infty$, and $\theta$, $\phi$ are co-ordinates on the two-sphere.  
The functions $\lambda$ and $\nu$ are some unspecified functions to start with. Such a Lorentzian manifold is called a BST provided there is a closed
orbit passing through each point in the interval $(r_1\,,\,r_2)$, which is stable under small perturbations of the initial conditions.

Starting from this definition, Perlick \cite{perlick} deduced that there can be two categories of BSTs. In this paper, we concentrate on 
one of these choices, called BST of type II given by:
\begin{eqnarray}
ds^2= -\frac{c^2 dt^2}{D + \alpha\sqrt{r^{-2}+K}} + \frac{dr^2}{\beta^2(1+Kr^2)}
+r^2(d\theta^2 + \sin^2 \theta\,d\phi^2)
\label{type2}
\end{eqnarray}
Here, the parameters $D$ and $K$ are real, $\alpha$ is real, and has to be positive as we show in sequel, and $\beta$ must be a positive rational number. 
Here $D$ and $\beta$ are dimensionless constants whereas $\alpha$ has the dimension of length and $K$ has the dimension of inverse squared length. 
In order to construct a phenomenological model of a galactic space-time using the metric of eq.(\ref{type2}), we minimally require three parameters, one fixing a
galactic scale, a second to parametrize the nature of closed orbits, and a third to fix the velocity in such an orbit. The work of
\cite{perlick} shows that the nature of the orbit is fixed by $\beta$, with $\beta = 1$ corresponding to Keplerian ellipses, deviations from
which can be obtained by tuning to different values of $\beta$. \footnote{As shown in \cite{perlick}, the ``apsidal angle'' of the trajectory, defined
as the azimuthal angle between a pericentric point and its successive apocentric point is given by $\frac{\pi}{\beta}$.}
We now have a choice, and can work with either $K=0$ or $D=0$.
The metric takes a simple form if we set $K = 0$ (the case $D=0$ will be briefly commented upon in a subsequent section) and takes the form
\begin{equation}
ds^2 = -\frac{c^2 dt^2}{D +\frac{\alpha}{r}} + \frac{dr^2}{\beta^2}
+r^2(d\theta^2 + \sin^2 \theta\,d\phi^2)\,.  
\label{type2a}
\end{equation}
As elaborated in the next section, the matter distribution that seeds the metric of eq.(\ref{type2a}) must necessarily have some exotic
components.  We will interpret this as dark matter. Our aim would thus be to model dark matter distribution in galaxies using the metric of
eq.(\ref{type2a}).  Most of our analysis in this section deals with the motion of a single massive object, moving in a time-like geodesic, 
in the background of the metric of eq.(\ref{type2a}). As in our model luminous matter moves in a space-time seeded by dark matter the contribution of 
luminous matter for the creation of BST is tacitly ignored. We expect our 
model to be maximally applicable in galaxies in which most of the matter content is dark matter, for example in low surface brightness (LSB) galaxies.

What lies at the centre of a galaxy is important. In our case, a simple computation shows that the Ricci scalar ${\mathcal R} \sim
r^{-2}$ and the Kretschmann scalar ${\mathcal K} \sim r^{-4}$ for small $r$. There is thus a genuine singularity at $r=0$ which is not
covered by a horizon, i.e the singularity is naked. Although it is generally believed that the centre of every galaxy contains a black
hole, there is no confirmed wisdom on the issue, and we can proceed with the fact that we have a naked singularity at the centre of our
galaxy. Note also that according to the classification of \cite{ve}, the BST in question is a strongly naked singularity, i.e it is not
covered by a photon sphere (where the bending angle of light becomes unbounded, or, alternatively, the conserved energy and angular
momentum for time-like geodesics diverge). In fact, using eq.(11) of \cite{ve}, it is easy to check that there is no physical solution for
the radius of a photon sphere in this case.

Coming back to the metric of eq.(\ref{type2a}), we will show in sequel that the parameters $\alpha$ and $D$ can be used to fix the scale of
the galaxy and the galaxy rotation curve \footnote{This is the reason the time coordinate is not scaled in eq.(\ref{type2a}) Our
calculations, are thus performed in a particular observer's frame, and should be taken in the spirit of a phenomenological model.} (for recent work on 
the rotation curves of disc galaxies and their dark matter distribution, from a different perspective, see, e.g. \cite{salucci}).
The constant $D$ will be taken to be positive, in order to preserve a Lorentzian signature at spatial infinity \footnote{Note that at
spatial infinity, constant time slices are spaces with conical defects, with a deficit angle related to $\beta$ (for $\beta < 1$, which
is the case in this paper). Our model will
however maximally hold till the galactic scale, which we will be determined by $r_s = \alpha/D$.}.  More over as the metric of
eq.(\ref{type2a}) is spherically symmetric and we can use this spherical symmetry of the system and set $\theta = \pi/2$, without loss of
generality. This will be assumed henceforth. 
Although as $r\to \infty$ the metric approaches the Minkowski form (with a suitable rescaling of the time coordinate), if one sets $\beta
\to 1$, one must be careful in tackling this limit. For $\beta = 1$, the energy density of the fluid, which seeds the BST, vanishes identically. In
fact it will be shown later that for a positive definite energy density BST, we require $0<\beta<1$.
\begin{figure}[t!]
\centering
\includegraphics{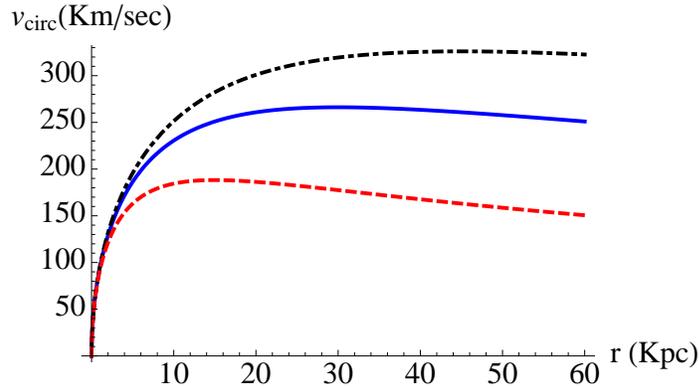}
\caption{Circular velocity (Km/sec) vs radial distance $r$ (Kpc) as predicted from eq.(\ref{vel}). The dashed red, solid blue, and 
dot-dashed black curves correspond to $D = 3 \times 10^5$, $1.5 \times 10^5$ and $10^5$ respectively. For all the curves, $\alpha =
  4.5 \times 10^6~{\rm Kpc}$.}
\label{vcirc}
\end{figure}

In the space-time specified by Eq.~(\ref{type2a}) one can find the geodesic equation, corresponding to the radial coordinate, for time-like geodesics as
\begin{eqnarray}
\ddot{r} + \frac{c^2\beta^2 \alpha}{2(Dr+\alpha)^2}\dot{t}^2
- r\beta^2 \dot{\phi}^2=0\,
\nonumber
\end{eqnarray}
where the dot denotes a derivative with respect to an affine parameter. For a circular orbit, this leads to
\begin{eqnarray}
v_{\rm circ}(r) = r\left(\frac{d\phi}{dt}\right) = \frac{c\sqrt{\alpha r}}{\sqrt{2}(Dr+\alpha)}\,
\label{vel}
\end{eqnarray}
Stability of circular orbits in the BST can be deduced directly from the work of \cite{perlick}, or from the fact that for time-like 
geodesic equations, we have, in terms of an effective potential $V(r)$,
\begin{equation}
{\dot r}^2 + V\left(r\right) =0,~~~~~V\left(r\right) = 
\beta^2\left(c^2 + \frac{h^2}{r^2}\right) - \frac{\beta^2\Gamma^2}{c^2}
\left(D + \frac{\alpha}{r}\right)
\end{equation}
where $h$ and $\Gamma$ are conserved quantities, the total angular momentum per unit rest mass and total energy per unit rest mass respectively, if the
affine parameter is chosen to be the proper time. 
For circular orbits, setting $V = 0$ and $\frac{dV}{dr}\vert_{r=r_0} = 0$, one can solve for $h$ and $\Gamma$ in terms of the (fixed) orbit radius,
$r_0$. Using these expressions, one can check that
\begin{equation}
\frac{d^2V}{dr^2}\vert_{r=r_0} = \frac{2\alpha \beta^2 c^2}
{r_0^2\left(\alpha + 2Dr_0\right)}
\end{equation}
which is positive definite for all $r_0$, indicating stability of the orbit. 

Equation (\ref{vel}) is graphically depicted in fig.(\ref{vcirc}). In this figure, we have chosen $\alpha = 4.5\times10^6~{\rm Kpc}$ and
$\beta = 0.8$, and the dashed red, solid blue, and dot-dashed black lines correspond to $D = 3 \times 10^5$, $1.5 \times 10^5$ and
$10^5$. It should be clear to the reader that by suitably varying the parameters $D$ and $\alpha$, $v_{\rm circ}$ can be kept
reasonably constant over a large range of $r$, and hence eq.(\ref{vel}) can be used to model circular velocity curves in
galaxies. For future reference, we record one set of parameters which we will use to highlight some features of our analysis :
\begin{eqnarray}
\alpha = 4.5\times10^6~{\rm Kpc},~~~D = 1.5\times 10^5,~~~\beta=0.8
\label{params}
\end{eqnarray}

One should be careful in comparing  eq.(\ref{vel}) with the circular
velocity of stars observed in galaxies. More appropriately,
eq.(\ref{vel}) should be thought of as the azimuthal velocity, and the mean azimuthal velocity becomes equal to the circular velocity (obtained
as a derivative of the galactic potential) only in the approximation that the velocity dispersions can be neglected \cite{binneytremaine}. 
In our case, what we simply have is a single luminous object moving in a closed orbit in a matter medium that seeds a BST, and hence there 
is no distribution of luminous matter. We also assume that the motion of this object in a time-like geodesic is sufficiently close to a circular trajectory. 
For recent related discussions, see \cite{mbcms}, \cite{bt}.

Now, from eq.(\ref{vel}) one can find the radius for which $v_{\rm circ}(r)$ maximizes and the maximum value of $v_{\rm circ}(r)$ as:
\begin{equation}
r_s = \frac{\alpha}{D},~~~~v_{\rm circ}^{\rm max} = \frac{c}{2\sqrt{2}}
\frac{1}{\sqrt{D}}
\end{equation} 
The value of $D$ and $\alpha$ can thus be estimated in principle by comparison with data for $v_{\rm circ}^{\rm max}$ and the radial distance 
at which the circular velocity maximizes.
\begin{figure}[t!]
\begin{minipage}[b]{0.5\linewidth}
\centering
\includegraphics[width=2.8in,height=2.3in]{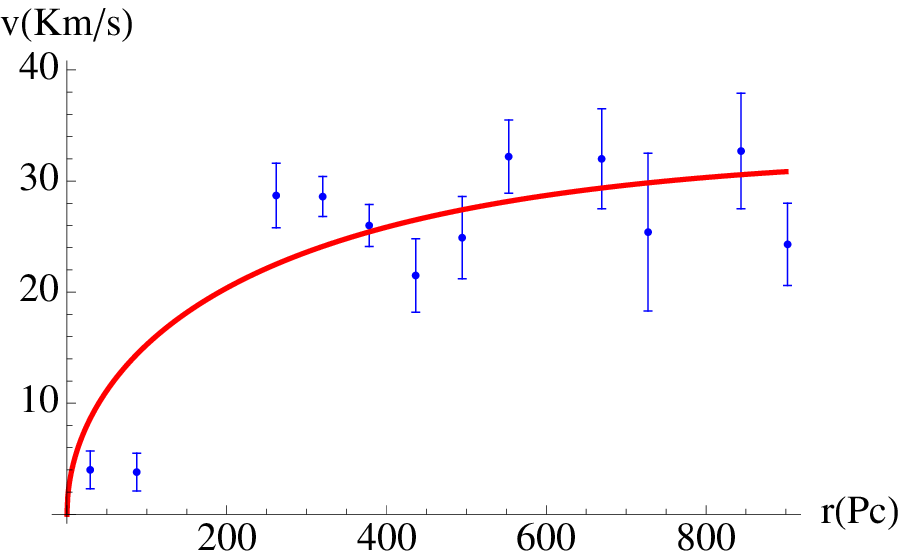}
\caption{Circular velocity of galaxy NGC4395 (blue dots) compared with the 
prediction of eq.(\ref{vel}) (red line) with $\alpha/D = 1545.45~{\rm Pc}$}
\label{4395}
\end{minipage}
\hspace{0.2cm}
\begin{minipage}[b]{0.5\linewidth}
\centering
\includegraphics[width=2.8in,height=2.3in]{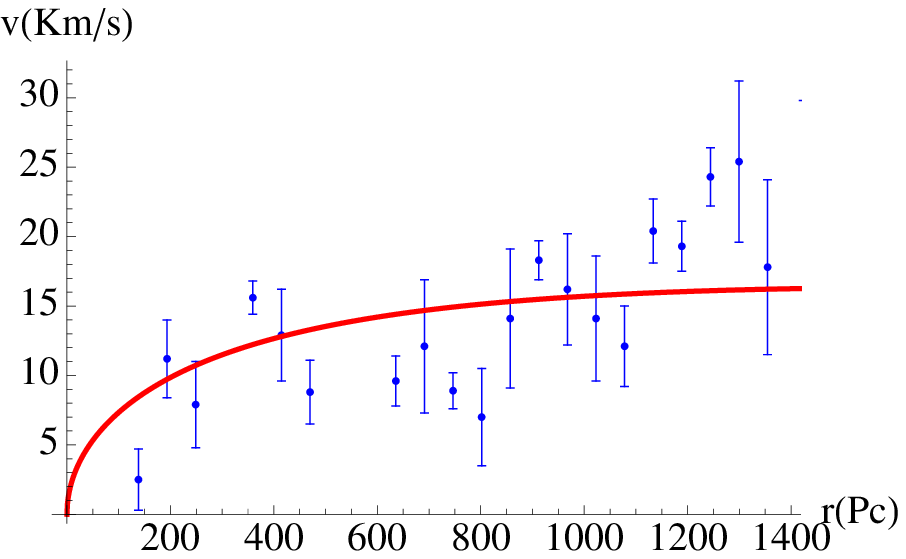}
\caption{Circular velocity of galaxy UGC1281 (blue dots) compared with 
the prediction of eq.(\ref{vel}) (red line) with $\alpha/D = 1785.71~{\rm Pc}$}
\label{1281}
\end{minipage}
\end{figure}
As we have noted before, our model is maximally suited for the study of galaxies that are dominated by dark matter, for example LSB galaxies. 
In fact, we find that by appropriately choosing $\alpha$ and $D$, the initial rise of the rotation curve, from eq.(\ref{vel}), is similar to galaxies of the LSB type. 
As concrete examples, in figs.(\ref{4395}) and (\ref{1281}), we have compared data (blue dots) from the galaxies NGC4395 and UGC1281 \cite{lsbdata} upto
$900$ and $1400$ Pc respectively, with the circular velocity of eq.(\ref{vel}). Here, $\alpha/D$ has been taken to be $1545.45$ and $1785.71$ Pc respectively. 

We note here that the derivation of eq.(\ref{vel}) was done from a GR perspective.  However, from a Newtonian point of view, such a velocity
distribution would involve a density profile ($G_N$ being Newton's constant) given by : \footnote{Here, the subscript $N$ denotes that we are working in the 
Newtonian approximation.} 
\begin{eqnarray}
\rho_N\left(r\right) &=& \frac{1}{4\pi G_Nr^2}\frac{d}{dr}\left(v_{\rm circ}^2r
\right) \nonumber\\
&=& \rho_{N0}\left[\left(\frac{r}{r_s}\right)^{-1}\left(1 + \frac{r}{r_s}
\right)^{-2} 
+ \left(1-\frac{r}{r_s}\right)\left(\frac{r}{r_s}\right)^{-1}\left(1 + 
\frac{r}{r_s}\right)^{-3} \right]\nonumber\\
&=&2\rho_{N0}\left(\frac{r}{r_s}\right)^{-1}\left(1 + \frac{r}{r_s}\right)^{-3}
\label{nfw}
\end{eqnarray}
where we have defined 
\begin{equation}
\rho_{N0} = \frac{c^2D}{\left(8\pi\alpha^2 G_N\right)}
\label{nfw2}
\end{equation}
In eq.(\ref{nfw}), the intermediate second line is kept to emphasize that near $r=r_s$, the second term drops out, and we recover the NFW
dark matter profile \cite{nfwprof}, while slightly away from this radius, this second term provides a correction to the NFW profile.  The third line of eq.(\ref{nfw}) indicates that 
in general, away from the flat region, a Hernquist \cite{hernquist} profile is predicted from our analysis, in the Newtonian limit. 
\footnote{To keep the discussion simple, and for consistency with the later sections, we will use $\alpha$ and $D$ as 
parameters in this section, rather than the standard ones used in the astronomy literature.} The appearance of the NFW profile is interesting. 
From eq.(\ref{vel}), we see that $dv_{\rm circ}/dr \propto (\alpha - Dr)$, and hence near $r = \alpha/D = r_s$, $v_{\rm circ}$ become independent of the radius. 
In this flat region, the NFW density profile is recovered, from eq.(\ref{nfw}). While this is not a derivation of the NFW model, we believe that our calculation 
provides a general relativistic justification as to why the NFW model provides a good fit for galaxy rotation curves in regions where they become flat. 

We note further that using eq.(\ref{nfw}), in the Newtonian approximation, we get the total dark matter mass of the galaxy as
\begin{equation}
M = \frac{\alpha c^2 R_{\rm max}^2}{2G_N \left(\alpha + DR_{\rm max}\right)^2}
\label{massraw}
\end{equation}
As an estimate, if we choose $R_{\rm max} = r_s = \alpha/D$, we obtain 
\begin{equation}
M = \frac{\alpha c^2}{8G_ND^2}
\label{mass}
\end{equation}
Specifically, choosing $\alpha = 1.7\times 10^{10}~{\rm Pc}$ and $D = 1.1\times10^7$ (choices made in the fit of fig.(\ref{4395})), 
we obtain an estimate of the dark matter halo mass of the galaxy NGC4395 as $\sim$ $4 \times 10^8 M_{\odot}$, while 
$\alpha = 7.5\times 10^{10}~{\rm Pc}$ and $D = 4.2\times10^7$ (choice made in the fit of fig.(\ref{1281})), 
gives an estimate of the dark matter halo mass for the galaxy UGC1281 as $\sim$ $1.2 \times 10^8 M_{\odot}$. As an aside, we make the following rough
estimate for the Milky Way. Assuming $v_{\rm circ}^{\rm max} \sim 265~{\rm Km/sec}$, we choose $\alpha = 2.7\times 10^6~{\rm Kpc}$ and
$D = 1.6\times 10^5$, so that $r_s \sim 16.9$ Kpc, roughly the radius of the Milky Way.  Then, the dark matter mass from eq.(\ref{mass}) is estimated 
as $\sim 3 \times 10^{11} M_{\odot}$, which is close to the lower bound predicted in \cite{battaglia}, using the NFW profile. Although our $v_{\rm circ}$ 
does not give a good fit in this case (possibly due to the relatively large amount of luminous matter in the Milky Way compared to LSB galaxies),
the prediction of the dark matter halo mass is close to observed results. 

We now briefly comment on the nature of radial geodesics on the equatorial plane. Using ${\dot \phi} = 0$, and ${\dot x}^{\mu}{\dot x}_{\mu} = 
-c^2$, where, as usual, ${\dot x}^{\mu}$ denotes the derivatives of the coordinates with respect to the proper time, we obtain
\begin{equation}
{\dot t} = \frac{\Gamma}{c^2}\left(D + \frac{\alpha}{r}\right),~~~{\dot r} 
= \beta\left[\frac{\Gamma^2}{c^2} \left(D + \frac{\alpha}{r}\right) - c^2\right]^{1 \over 2}
\label{geot}
\end{equation}
where $\Gamma$ as noted previously is the total energy of the particle per unit rest mass. Hence, the radial velocity is obtained as
\begin{equation}
v_{\rm rad} = \frac{\beta c \sqrt{r}}{\Gamma\left(\alpha + Dr\right)}
\left[\Gamma^2\left(Dr + \alpha\right) - c^4r\right]^{1 \over 2}
\label{geovrad}
\end{equation}
The radial velocity thus becomes zero for a maximum radius
\begin{eqnarray}
r_{\rm max} = \frac{\alpha \Gamma^2}{c^4 -   D \Gamma^2}
\label{rm}
\end{eqnarray}
hence, $r_{\rm max}$ can be finite depending on the value of $\Gamma$. For example, particles for
which $\Gamma = c^2/\sqrt{2D}$ have $r_{\rm max} = \alpha/D$. These particle are thus constrained to remain within the galaxy, up to the point where
$v_{\rm circ}$ maximizes. There is no such constraint on light-like particles, for which ${\dot x}^{\mu}{\dot x}_{\mu} = 0$, i.e any light
like particle will escape to asymptotic infinity.

\section{The Nature of Matter Sourcing BSTs}
\label{msource}
Having provided evidence that BSTs are relevant objects for modeling dark matter in galaxies, it is important to understand the nature of
matter seeding the BST. In fact, this is where we will find that the nature of such matter is exotic, i.e they can have negative
pressure. We will first present some considerations on the generic features of the energy-momentum tensor of BSTs. We then show that the
system can be described by a two-fluid model, and find its solution.

\subsection{The Energy-Momentum Tensor in a Class of BSTs}

In this subsection, we study the nature of matter that can seed the BSTs that we have discussed till now.  Starting from the metric of
eq.(\ref{type2a}), we can write the energy density and the principal pressures as:
\begin{eqnarray}
\rho(r) &=& -T^0_0 = \frac{1-\beta^2}{\kappa r^2}\,,
\label{rho}\\
p_r(r)  &=& T_1^1 = \frac{\beta^2(2\alpha + Dr)-(\alpha + Dr)}
{\kappa r^2(Dr + \alpha)}\,,
\label{pre}\\
p_\perp(r)  &=& T^2_2 = T^3_3 = \frac{\alpha\beta^2(\alpha - 2Dr)}
{4r^2\kappa(Dr+\alpha)^2}\,.
\label{pperp}
\end{eqnarray}
where we have assumed Einstein's equation $G_{\mu \nu}=\kappa T_{\mu \nu}$ to hold in BST of type II, with $\kappa = 8\pi G_N/c^4$. 
From the form of the energy density and pressures one can show that they satisfy the relation
\begin{eqnarray}
\frac{dp_r}{dr} = -(\rho + p_r)\frac{d\nu}{dr} + \frac2r(p_\perp - p_r)
\label{tov}
\end{eqnarray}
which can be interpreted as the Tolman-Oppenheimer-Volkoff equation for an anisotropic fluid in BST of type II. In this case the function $\nu$ can be obtained
from the form of the general metric in eq.(\ref{invl}) and the specific form of BST of type II as given in eq.(\ref{type2a}). Note that by assuming 
\begin{eqnarray}
0 < \beta < 1\,,
\label{bcond}
\end{eqnarray}
one can always maintain a positive energy density for this kind of space-time. But the above condition does not guarantee that $G_1^1$
and $G_2^2$ are positive definite and consequently there might be a situation where some of the principle pressures turns out to be
negative. We will come back to this shortly.

For the moment, we note that for a physically reasonable space-time, we require the Weak Energy Condition (WEC) to be valid.  Explicitly,
the WEC states that
\begin{equation}
\rho(r) \ge 0\,,\,\,\,\rho(r)+p_r(r)\ge 0\,,\,\,\,\rho(r)+p_\perp(r)
\ge 0\,,
\label{wec}
\end{equation}
While positivity of the energy density is guaranteed from eqs. (\ref{rho}) and (\ref{bcond}), it can be easily checked that
\begin{equation}
\rho + p_r = \frac{\alpha \beta^2}{\kappa r^2\left(\alpha +Dr\right)}
\end{equation}
which is positive for all positive values of $\alpha$ and $D$. The final condition of eq.(\ref{wec}) can be shown to yield 
\begin{equation}
\alpha^2\left(4 -3 \beta^2\right) + \alpha Dr\left(8 - 10\beta^2\right)  + 
4D^2r^2\left(1 - \beta^2\right)  > 0
\label{wec2a}
\end{equation}
which simplifies for a particular choice of $\beta$ (dictated by eq.(\ref{bcond})). We choose $\beta = 0.8$, for which eq.(\ref{wec2a})
is valid for all positive values of $\alpha$ and $D$. Hence, the WEC is satisfied for the BSTs that we consider in this paper.

The expression of the pressures specify that we cannot use an ideal fluid model for BST of type II as $p_r \neq p_\perp$. The anisotropy
in the pressure can be calculated from
\begin{equation}
p_r - p_{\perp} = \frac{0.12\left(\alpha^2 + 2\alpha D r - 
3D^2r^2\right)}{\kappa r^2\left(\alpha + Dr\right)^2},
\label{pdiff}
\end{equation}
where we have substituted $\beta = 0.8$. This has a zero at $r = \alpha/D$, where the anisotropy disappears, and for $r > \alpha/D$,
$p_r < p_{\perp}$. The pressure anisotropy dictates that if we want to model the matter content in the space-time by ideal fluids then one
can minimally employ two ideal fluids in relative motion, which can give rise to such an anisotropic pressure.  We now provide such a
two-fluid model solution for BSTs of type II.

\subsection{A Two-Fluid Model for BSTs and its Solution}

We begin by reviewing the basic formalism of \cite{letelier},\cite{bayin} for a two-fluid astrophysical model, which we wish to connect to the 
formulae obtained in the previous section. The energy-momentum tensor is composed of two noninteracting perfect fluids and its form is given as
\begin{eqnarray}
T_{\mu \nu} = (p_1 + \rho_1)u_\mu u_\nu + p_1 g_{\mu \nu} + (p_2 +
\rho_2)v_{\mu} v_\nu + p_2 g_{\mu \nu}\,,
\label{2femt}
\end{eqnarray}
where 
$$u^\mu u_\mu = v^\mu v_\mu = -1\,.$$ After a linear transformation in the 4-velocity space \cite{letelier}, the energy-momentum tensor can be
written in a standard form as
\begin{eqnarray}
T_{\mu \nu} = (\rho + p_\perp)w_\mu w_\nu + p_\perp g_{\mu \nu} +
(p_r - p_\perp) y_\mu y_\nu\,,
\label{2frmt1}
\end{eqnarray}
where 
$$w^\mu w_\mu =-1\,,\,\,\,y^\mu y_\mu=1\,,\,\,\,w^\mu y_\mu =0\,,$$ where $w_\mu$ represents the 4-velocity of the effective two-fluid
system and $y_\mu$ is a spacelike vector along the direction of anisotropy. In terms of the energy densities and pressures of the
component ideal fluids, denoted by $\rho_1, p_1$ and $\rho_2, p_2$ respectively, the energy density $\rho$ and pressure $p$ appearing in
eq.(\ref{2femt}) can be shown to be given by
\cite{letelier},\cite{bayin}
\begin{eqnarray}
\rho &=& -\frac12(\rho_1 - p_1 + \rho_2 - p_2) + \frac12 \left[
(p_1 + \rho_1 + p_2 + \rho_2)^2\right.\nonumber\\ 
&+&\left. 4(p_1 + \rho_1)(p_2 + \rho_2)
\left\{(u_\mu v^\mu)^2 -1\right\}\right]^{1/2}\,,
\label{rhot}\\
p_r &=&\frac12(\rho_1 - p_1 + \rho_2 - p_2) + \frac12 \left[
(p_1 + \rho_1 - p_2 - \rho_2)^2\right.\nonumber\\ 
&+&\left. 4(u_\mu v^\mu)^2(p_1 + \rho_1)(p_2 + \rho_2)\right]^{1/2}\,,
\label{prt}\\
p_\perp &=& p_1 + p_2\,.
\label{pperpt}
\end{eqnarray}
Now a specific choice of co-ordinates has to be made. For spherically symmetric anisotropic fluids one may choose
\begin{eqnarray}
y^0=y^2=y^3=0;~~~~~w^1=w^2=w^3=0
\end{eqnarray}
so that $y^1y_1= 1$ and $w^0w_0=-1$.With these choices, the energy momentum tensor becomes 
\begin{equation}
T^0_0 = -\rho,~~~T^1_1 = p_r,~~~T^2_2 = T^3_3 = p_{\perp}
\end{equation}
This form of the energy-momentum tensor is the one that we obtained
for BSTs of type II in eqs.(\ref{rho}), (\ref{pre}) and (\ref{pperp}).
Now, the Einstein's equation gives
\begin{eqnarray}
-\frac{1}{\kappa}G_0^0 = \rho,~~~ \frac{1}{\kappa}G_1^1 = p_r, ~~~
\frac{1}{\kappa}G_2^2 = p_1 + p_2
\label{main2fluid}
\end{eqnarray}
where $G_0^0$, $G_1^1$ and $G_2^2$ can be obtained from eqs.(\ref{rho}), (\ref{pre}) and (\ref{pperp}), respectively, and
$\rho$, $p_r$ and $p_{\perp}$ are given by eqs.(\ref{rhot}), (\ref{prt}) and (\ref{pperpt}) respectively.  Calling $u_\mu v^\mu =
K$, we have three equations for five unknowns: $$\rho_1(r)\,,\,\rho_2(r)\,,\,p_1(r)\,,\,p_2(r)\,,\,K(r)$$ defining
the two-fluid BST of type II. The system of equations can yield a solution if one assumes the simple barotropic equations of state for
the two fluids
\begin{eqnarray}
p_1 = \gamma_1 \rho_1\,,\,\,\,\,\,p_2=\gamma_2 \rho_2
\label{eos2}
\end{eqnarray}
where $\gamma_1$ and $\gamma_2$ are constants. In this article we will restrict the possible values of these parameters by
\begin{eqnarray}
-1 < \gamma_1\,,\,\gamma_2 < 1
\label{gamval}
\end{eqnarray}
The above equation can be interpreted as a restriction on the nature of the fluids acting as source of BST of type II. The lower limit of
the state parameters specify that we are neglecting phantom fields\footnote{One can easily show that if either $\gamma_{1}$ (or 
$\gamma_2)=-1$, then the anisotropy in the pressure vanishes and so in our case none of the state parameters can be negative unity.}  from
the matter part and the upper limit is set such that the equations of state does not become too steep.  

Now, we present the solution for the two-fluid model. This is done by using eq.(\ref{eos2}) and eqs.(\ref{rhot}) - (\ref{pperpt}),
and after some algebra, we get the following simple expressions \footnote{These formulae can be easily checked by the reader, by
explicit substitution from eq.(\ref{main2fluid}), and eqs.(\ref{rhot}) - (\ref{pperpt}).}:
\begin{eqnarray}
\rho_1(r) &=& \frac{\gamma_2\left(G_2^2 - G_1^1 - G_0^0\right)-G_2^2}
{\kappa(\gamma_2 - \gamma_1)}\,,
\label{r1f}\\
\rho_2(r) &=& \frac{\gamma_1(G_0^0+G_1^1-G_2^2) + G_2^2}{\kappa
(\gamma_2 -\gamma_1)}\,,
\label{r2f}\\
u_{\mu}v^{\mu} \equiv K &=& 
-\left[\frac{\left[G_1^1 + \kappa(\rho_1 - \gamma_2\rho_2)\right]
\left[G_1^1 + \kappa(\rho_2 - \gamma_1\rho_1)\right]}
{\kappa^2 \rho_1 \rho_2 (1+\gamma_1)(1+\gamma_2)}\right]^{1/2}
\label{umuvmu}
\end{eqnarray}
The negative sign in the last equation is chosen because $u$ and $v$ are assumed to be time-like vectors.  We now analyze these equations
in detail. First, we present the solutions for $\rho_1$ and $\rho_2$ in terms of the BST parameters.  It can be shown that
\begin{equation}
\rho_1 = \frac{{\mathcal A}_1\left(r\right) + {\mathcal C}_1}{{\mathcal D}
\left(r\right)},~~~
\rho_2 = \frac{{\mathcal A}_2\left(r\right) + {\mathcal C}_2}{{\mathcal D}
\left(r\right)},
\end{equation}
where ${\mathcal C}_1$ and ${\mathcal C}_2$ are constants in the sense that they do not depend on $r$, and we obtain
\begin{eqnarray}
{\mathcal A}_1\left(r\right) &=& 
-8\gamma_2 r^2D^2\left(1 - \beta^2\right) + 2\alpha Dr\left[\beta^2
\left(11\gamma_2 - 1\right) - 8\gamma_2\right]\,,
\label{a1}\\
{\mathcal A}_2\left(r\right) &=& 
8\gamma_1 r^2D^2\left(1 - \beta^2\right) + 2\alpha Dr\left[\beta^2
\left(11\gamma_1 - 1\right) - 8\gamma_1\right]\,,
\label{a2}\\
{\mathcal C}_1 &=& \alpha^2\left[\beta^2\left(11\gamma_2 + 1\right) - 
8\gamma_2\right]\,,
\label{c1}\\
{\mathcal C}_2 &=& -\alpha^2\left[\beta^2\left(11\gamma_1 + 1\right) - 
8\gamma_1\right]\,,
\label{c2}\\
{\mathcal D}\left(r\right) &=& 4r^2\kappa \left(\alpha + Dr\right)^2
\left(\gamma_1 - \gamma_2\right)\,.
\label{d1}
\end{eqnarray}
\begin{figure}[t!]
\begin{minipage}[b]{0.5\linewidth}
\centering
\includegraphics[width=2.8in,height=2.3in]{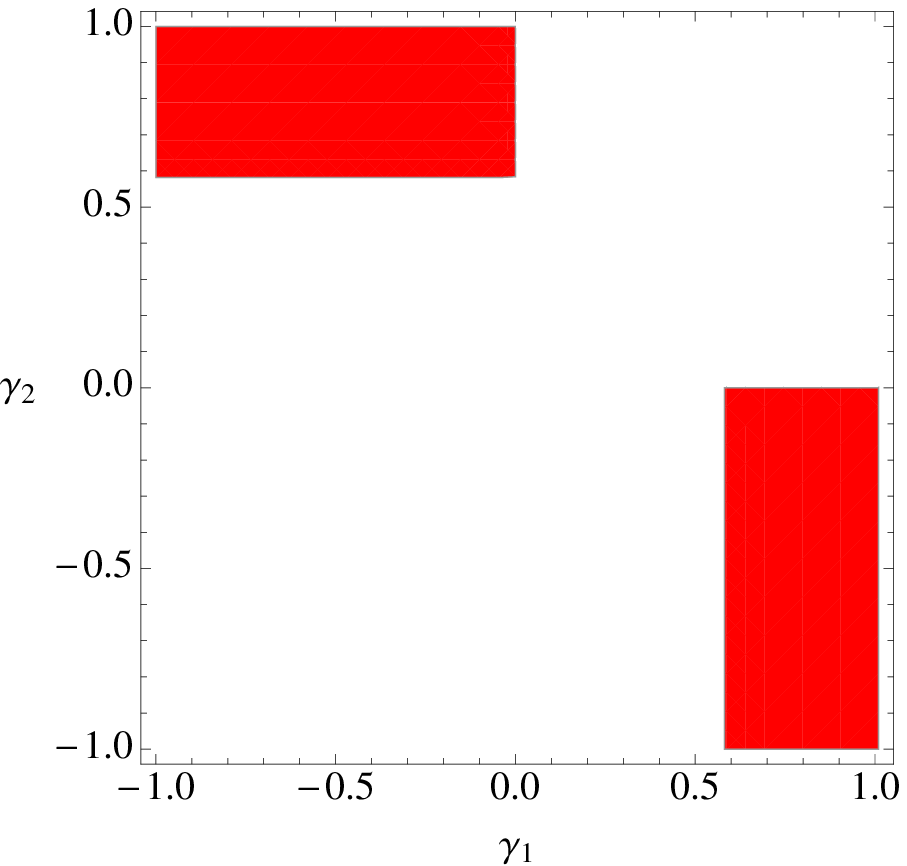}
\caption{Admissible ranges of $\gamma_1$ and $\gamma_2$, from positivity 
of $\rho_1$ and $\rho_2$.}
\label{a1a2rho}
\end{minipage}
\hspace{0.2cm}
\begin{minipage}[b]{0.5\linewidth}
\centering
\includegraphics[width=2.8in,height=2.3in]{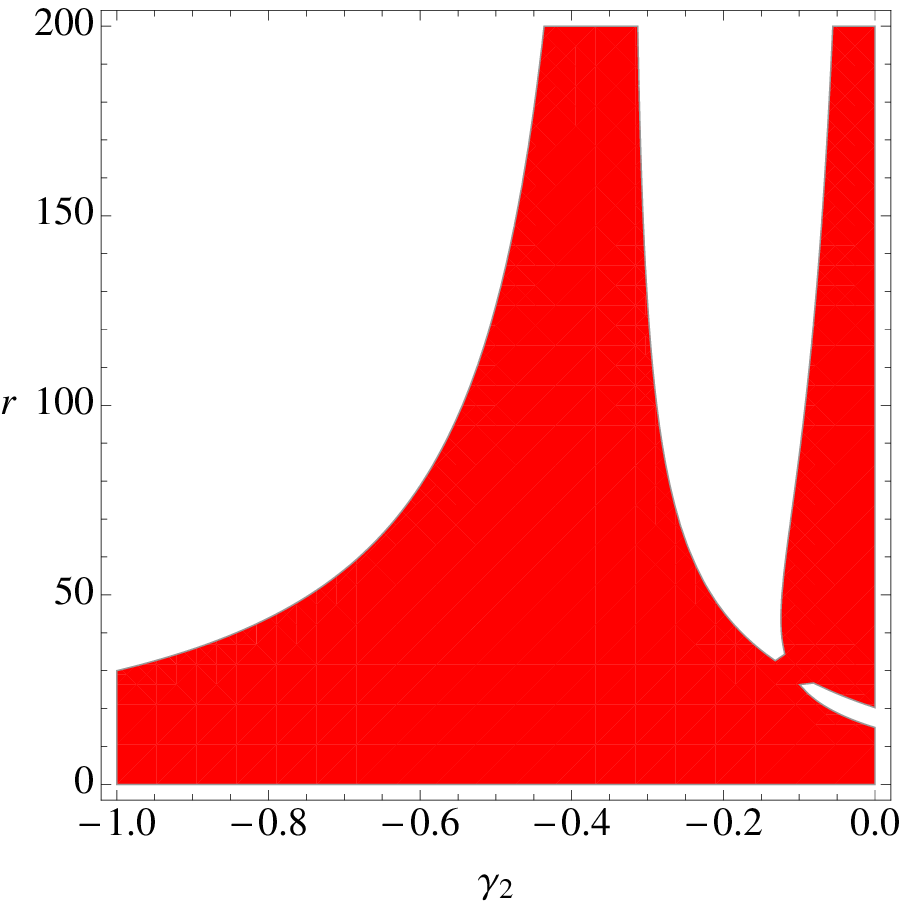}
\caption{Admissible range for $r$ and $\gamma_2$, from reality of 
$u_{\mu}v^{\mu}$, with $\gamma_1 = 0.9$}
\label{a1a2k}
\end{minipage}
\end{figure}
For physically meaningful solutions we require the energy densities to be positive throughout. First, note that $\gamma_1 \neq \gamma_2$ for
finiteness of the energy densities. Next, note that either of $\gamma_1$ or $\gamma_2$ cannot be zero. This is because we find that
\begin{eqnarray}
\rho_1 &=& \frac{\alpha\beta^2\left(\alpha - 2Dr\right)}
{4r^2\kappa\gamma_1\left(\alpha + Dr\right)^2},~~~{\rm for}~~\gamma_2 = 0 \\
\rho_2 &=& \frac{\alpha\beta^2\left(\alpha - 2Dr\right)}{4r^2\kappa\gamma_2
\left(\alpha + Dr\right)^2},~~~{\rm for}~~\gamma_1 = 0 
\end{eqnarray}
that become negative for $r > \alpha/(2D)$. We thus require $\gamma_1\ne 0,~\gamma_2\ne 0$. This is important, as we have
established that neither of the two perfect fluids producing the anisotropic fluid responsible for BST of type II can be ideally
dust. Both must have non-zero pressure.  Now, for large $r$, from eqs.(\ref{a1}) and (\ref{a2}), we see that the $r^2$ dependent terms
will dominate in the numerators of $\rho_1$ and $\rho_2$.  It can be seen that $\gamma_1$ and $\gamma_2$ have to be of opposite signs for
positiveness of both $\rho_1$ and $\rho_2$, in this limit.  For small $r$, we can neglect the terms ${\mathcal A}_1$ and ${\mathcal A}_2$ in
the numerators of $\rho_1$ and $\rho_2$. In both cases, there is an allowed range for $\gamma_1$ and $\gamma_2$ for which the energy densities
are positive. As an illustration, using the values of the parameters as given in eq.(\ref{params}), and setting $r = 1~{\rm Kpc}$, 
the admissible range of $\gamma_1$ and $\gamma_2$ (where both $\rho_1$ and $\rho_2$ are positive) is shown in the shaded region of fig.(\ref{a1a2rho}).

Note also that $\gamma_1$ and $\gamma_2$ might be further constrained, from the condition that in eq.(\ref{umuvmu}), $K$ should
be real (and negative). Here, we will present a graphical analysis as the expression for $K$ turns out to be complicated. Setting the BST
parameters of eq.(\ref{params}), and further setting $\gamma_1 = 0.9$, we find that $K$ is real in the shaded region depicted in fig.(\ref{a1a2k}). 
It should be noted here that as mentioned at the end of subsection 3.1, for $\beta = 0.8$, the anisotropy term in eq.(\ref{2frmt1}) disappears 
for $r = \alpha/D = 30~{\rm Kpc}$ (using the parameters of eq.(\ref{params})), and our two fluid model ceases to be valid beyond this radius.
\begin{figure}[t!]
\begin{minipage}[b]{0.5\linewidth}
\centering
\includegraphics[width=2.8in,height=2.3in]{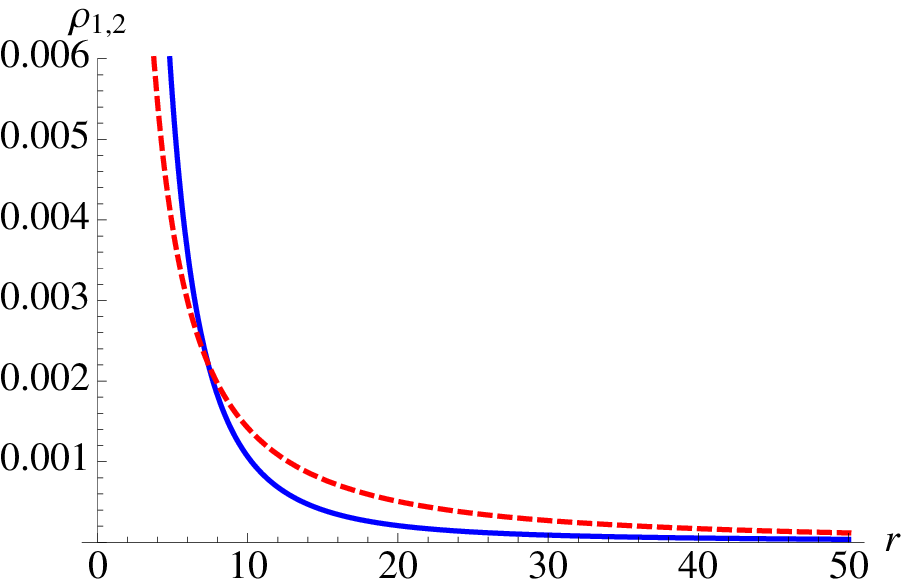}
\caption{Plots of $\rho_1$ (solid blue) and $\rho_2$ (dashed red) vs 
$r$ for $\gamma_1 = 0.9$, $\gamma_2 = -0.4$. See text for more details.}
\label{rhovsr}
\end{minipage}
\hspace{0.2cm}
\begin{minipage}[b]{0.5\linewidth}
\centering
\includegraphics[width=2.8in,height=2.3in]{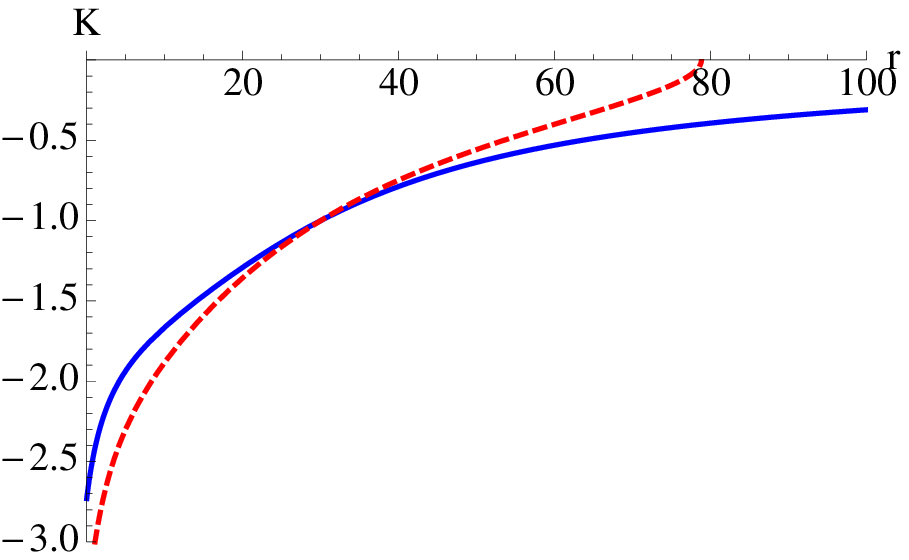}
\caption{$K$ vs $r$ for $\gamma_1 = 0.9$, $\gamma_2 = -0.4$ 
(solid blue) and $\gamma_1 = 0.9$, $\gamma_2 = -0.6$ (dashed red). 
Details in text.}
\label{kvsr}
\end{minipage}
\end{figure}

\begin{figure}[t!]
\begin{minipage}[b]{0.5\linewidth}
\centering
\includegraphics[width=2.8in,height=2.3in]{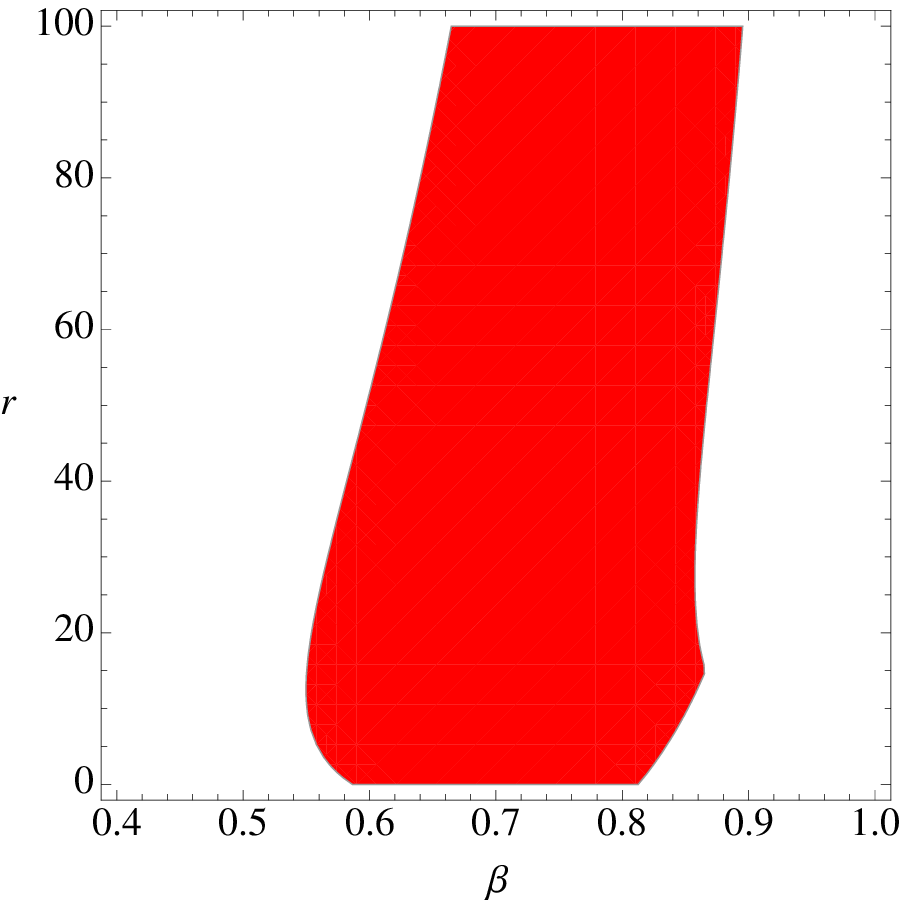}
\caption{Admissible range for $\beta$, for $\gamma_1 = 0.9$, $\gamma_2=-0.4$ See text for more details.}
\label{rvsbeta1}
\end{minipage}
\hspace{0.2cm}
\begin{minipage}[b]{0.5\linewidth}
\centering
\includegraphics[width=2.8in,height=2.3in]{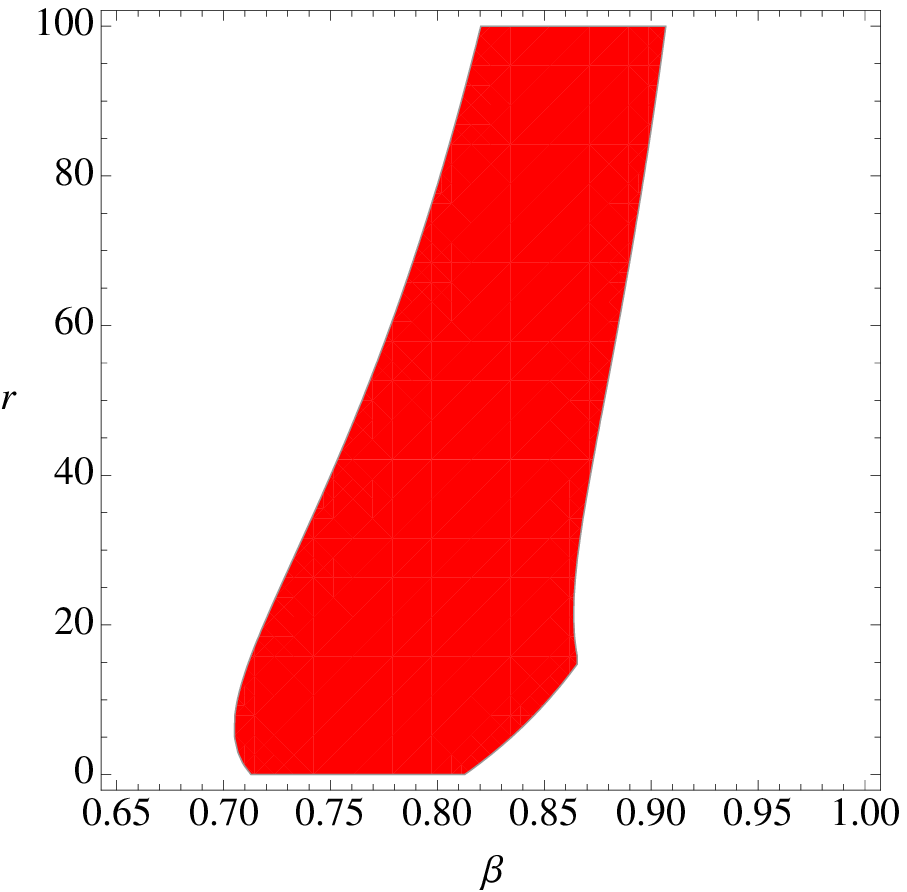}
\caption{Admissible range for $\beta$, for $\gamma_1 = 0.9$, $\gamma_2=-0.6$ See text for more details.}
\label{rvsbeta2}
\end{minipage}
\end{figure}

To illustrate the preceding discussion, we have plotted, in fig.(\ref{rhovsr}), $\rho_1$ (solid blue line) and $\rho_2$ (dashed
red line) vs $r$ for $\gamma_1 = 0.9$, $\gamma_2 = -0.4$. These are seen to be positive definite, in line with figs.(\ref{a1a2rho}) and
(\ref{a1a2k}).  The BST parameters have been chosen to be the same as in the previous discussion. With these parameters, fig.(\ref{kvsr})
shows the variation of $K$ (of eq.(\ref{umuvmu})) with $r$, for $\gamma_1 = 0.9$, $\gamma_2 = -0.4$ (solid blue line) and $\gamma_1 = 0.9$,
$\gamma_2 = -0.6$ (dashed red line).  At $r = 30~{\rm Kpc}$, at which our two fluid model ceases to be valid, $K=-1$ for 
any choice of $\gamma_1$ and $\gamma_2$. Finally, we point out that although we have used $\beta = 0.8$ throughout this paper, it is easy to 
derive bounds on $\beta$ for given values of the other parameters. Taking $\alpha = 4.5\times 10^6{\rm Kpc}$, $D = 1.5 \times 10^5$, in fig.(\ref{rvsbeta1}) and 
(\ref{rvsbeta2}), we have plotted, for $\gamma_1 = 0.9$, $\gamma_2 = -0.4$, and $\gamma_1 = 0.9$, $\gamma_2 = -0.6$ respectively, the 
allowed range for $\beta$, for $\rho_1$, $\rho_2$ (of eqs.(\ref{r1f}), (\ref{r2f})) to be real and positive, and $K$ of eq.(\ref{umuvmu})) to be real and negative. 
For the range of $r$ that we are interested in, $\beta=0.8$ is indeed seen to be a valid choice. 

Having presented our main results, we now offer some general comments that are relevant to our previous discussion. 
\section{General Comments and Discussions}

We will start this section by contrasting our approach with some existing ones in the literature. 
As is well known, application of Newtonian or semi-Newtonian physics to understand the properties of galaxies is standard practice. 
The main reason for such an approach may be due to the paradigm in which one tries to frame the dynamics of stellar motion in galaxies. We are
aware of some earlier attempts \cite{mdroberts},\cite{Harko:2011nu}, \cite{Bharadwaj:2003iw}, where a general relativistic approach was used to
understand galactic physics. In these works, the authors try to model the galactic space-time with an external Schwarzschild space-time. 
In a general relativistic perlance one has to admit that dark matter which produces the galactic space-time can have
pressure. However, in the conventional set up one generally works with the energy density of dark matter, and its pressure is set to zero, although
the authors of \cite{Dalcanton:2010bp} try to model a galaxy with matter which has pressure. In a different context, \cite{Bucher:1998mh} states 
the possibility of negative pressure dark matter in cosmology. But such unconventional analysis is relatively rare in the literature.  In \cite{Bharadwaj:2003iw}, 
the authors note that in modeling a galactic space-time, dark matter can have non-trivial pressure.  In \cite{Harko:2011nu} there was an attempt to produce 
a galactic space-time using a two-fluid model. In this regard, we observe in this paper, we have not used any form of a space-time that is motivated by 
an exterior Schwarzschild solution. In our case, the BST of type II does not asymptote to a Schwarzschild solution in any region. 
It is a remarkable property that using BSTs, we can get reasonable predictions for quantities that are of interest to astronomers, in particular popular
dark matter density profiles. 
The BST background in this paper is produced by an anisotropic fluid, which can be modelled by two perfect fluids which are moving bodily with respect to each 
other in the frame where the calculations are performed. The two-fluid model itself suggests that there can be turbulent motions of the perfect (dark matter) fluids
inside the galaxy. Due to the particular nature of the anisotropic pressures, one cannot extend the two-fluid model far away from the galactic center, as we
have noted. With the BST parameters specified in eq.(\ref{params}), at around $30 {\rm Kpc}$ the anisotropy in pressure vanishes and consequently the two-fluid
picture ceases to be valid. For phenomenological purpose it is assumed that all the local structures of a typical galaxy modelled by BST of type II
ends around $30 {\rm Kpc}$. Admittedly, the dark matter distribution will be present beyond the radial cut-off that we have discussed. But by suitably choosing 
parameters, as discussed after eq.(\ref{rm}), it is possible to ensure that no luminous matter goes outside the cut-off, except light. A similar problem was 
addressed in \cite{Nucamendi:2000jw}, where the authors tried to model the source of the galactic space-time by scalar fields, and attempted to fit their galactic 
solution to an exterior Schwarzschild space-time. We will comment on this feature in a while. 

Also, our BST has a naked singularity at $r=0$. Whether in nature one can find a naked singularity is still a debatable issue, due to the cosmic censorship hypothesis. 
But it is also known that general general relativistic effects can produce naked singularities \cite{ve}, \cite{Sahu:2012er}, \cite{joshi}. In light of these
arguments, we assume that the naked singularity present at the center of the galaxy in a BST of type II could have been produced by some gravitational collapse of 
dark matter dominated overdensed regions in the early universe.

Note further that if one employes a two-fluid model in the interior of the galaxy none of the two perfect fluids can be
dust. This directly implies that the two component dark matter employed must have pressure. Our observation here is that 
one component of dark matter must have negative pressure. If we did not use the two-fluid model, and worked with an anisotropic fluid instead, 
then also it is seen that pressures turns out to be negative at around $30 {\rm Kpc}$. Thus, our model predicts that there can be dark matter candidates 
with negative pressure.

Now, we present a few results that we have not discussed in detail till now. In particular, we will comment on the BST of eq.(\ref{type2}) with $D=0$ (this metric also has 
two free parameters that can be used for galactic modeling). We also present the construction of a new space-time that is internally a
BST and seeds an exterior Schwarzschild solution.

\subsection{Analysis of an Alternative BST Metric}

We start from eq.(\ref{type2}) and set $D=0$, so that our metric is now given by
\begin{eqnarray}
ds^2= -\frac{c^2 dt^2}{\alpha\sqrt{r^{-2}+K}} +
\frac{dr^2}{\beta^2(1+Kr^2)} +r^2(d\theta^2 + \sin^2
\theta\,d\phi^2)\,.
\label{type2b}
\end{eqnarray}
$K$ has the dimensions of inverse square length, and $\alpha$ has dimensions of length, as usual. In this case, in the $\theta = \pi/2$
plane, we obtain the circular velocity
\begin{equation}
v_{\rm circ}= \frac{c\sqrt{r}}{\sqrt{2\alpha}\left(1 + Kr^2\right)^{3 \over 4}}
\end{equation}
The maximum velocity occurs for $r = 1/\sqrt{2K}$, and at this radius, we obtain
\begin{equation}
v_{\rm circ}^{\rm max} = \frac{c}{\left(27 K\right)^{1 \over 4}\sqrt{\alpha}}
\end{equation}
In the Newtonian limit, near $r = 1/\sqrt{2K}$, we obtain the mass density profile
\begin{equation}
\rho_N^1 = \frac{\rho_{N0}}{r\left(1 + Kr^2\right)^{3 \over 2}},~~~~~\rho_{N0} = 
\frac{c^2}{8\pi\alpha G_N}
\end{equation}
while in general, away from this radius, we obtain 
\begin{equation}
\rho_N = \rho_{N0}\frac{\left(2-Kr^2\right)}{r\left(1 + Kr^2\right)^{3 \over 2}}
\end{equation}
The metric of eq.(\ref{type2b}) is thus meaningful only till $r = \sqrt{2/K}$, i.e twice the radius at which the circular velocity maximizes. 
This is interesting, and our preliminary checks indicate that the metric of eq.(\ref{type2b}) can also be effectively used to model galactic 
regions, by fitting $\alpha$ and $K$. We will, however, defer a full solution to a future publication.

\subsection{An External Schwarzschild Solution}

It is interesting to extend the analysis of the previous sections to a case where an internal BST sources an external Schwarzschild
solution (recent work in this direction for a class of naked singularities that may form in gravitational collapse appear in \cite{joshi}). 
This might be a more realistic scenario to compare observational data on gravitational lensing. In \cite{perlick}, an asymptotically
Minkowskian BST was constructed with $\beta=1$, but it was shown that this violates the weak energy condition at spatial infinity. An external
Schwarzschild metric seeded by an internal BST will not have this problem. We will now present
such a solution in our case, where the internal BST sources the external Schwarzschild solution :
\begin{eqnarray}
ds_i^2 &=& -\frac{3\beta^2}{\left(1 + \frac{2r_B}{r}\right)}c^2dt^2 + 
\frac{1}{\beta^2}dr^2 + r^2d\Omega^2 \nonumber\\
ds_e^2 &=& -\left(1 - \frac{2M r_B}{r}\right)c^2dt^2 + 
\frac{dr^2}{\left(1 - \frac{2M r_B}{r}\right)} + r^2d\Omega^2
\label{mixed}
\end{eqnarray}
where $M$ is a dimensionless constant and we have suppressed the Newton's constant in the external solution. 
Here, we require the identification $\beta^2 = \left(1 - 2M\right)$ and $r_B = \alpha/(2D)$, where $\beta$,
$\alpha$ and $D$ are the same parameters that appear in eq.(\ref{type2a}). The matching radius is at $r = r_B$, where the radial and transverse
pressures vanish, for $\beta^2 = 3/5$. The Schwarzschild radius is then seen to be at $2r_B/5$. The attractive feature of this solution is that it is 
asymptotically flat, and hence conventional computations of gravitational lensing can be carried out here. Interestingly, we note that $\beta$ is constrained
to be irrational in this case,  indicating that a circular orbit may not be closed under perturbation. The structure of the space-time of eq.(\ref{mixed}) 
requires extensive analysis, and we expect to report on this shortly \cite{dbs}.

\section{Conclusions and Future Directions}

In this paper, we have attempted to model some aspects of the physics of galaxies, by general relativistic arguments, using a Bertrand space-time. 
It is seen that some important properties of typical low surface brightness galaxies, such as the circular velocity and mass density profiles, can be obtained 
from our analysis. We are led to the conclusion that the matter which seeds our BST  is an anisotropic dark matter fluid. Under some specific choice of parameters 
in the metric of our BST, we can model this anisotropic fluid by two perfect dark fluids. The novel feature of our approach is that we did not assume a prior
knowledge of the circular velocities, and further, the galactic space-time was not modeled from a specific Schwarzschild solution. We have also
presented some preliminary results on a new space-time that is internally a BST and externally Schwarzschild.
As we have noted, there is a singularity at the center of the galaxy which is inevitable if we apply BST to model galactic properties. As
all the indications about the mass density profile seeding the BST coincides with dark matter density profiles, we may conclude that the central naked 
singularity was caused due to gravitational collapse of an overdensed region of dark matter in some earlier epoch. 

The main drawback of our model is that we are unable to predict the end of galactic matter using our two fluid model.  It is seen that the important properties 
of the galactic structure can be contained inside a radial cut-off, but the space-time extends beyond this. One can interpret this result as follows. Local properties of a 
typical galaxy can be contained within the radial cut-off and no ordinary matter can leave this local gravitational system, except light. The only matter which
extends outside the cut-off must be dark matter remnants outside the main galactic structure. The galactic model employed in this paper predicts that there will be 
some dark matter distribution outside galaxies. We believe that calculations of gravitational lensing might put our model on a firmer footing. Work is in progress
in this direction.

Finally, in this paper, most of our calculations are based on a single massive object moving in a time-like geodesic in a BST. It will be interesting to numerically 
simulate the dynamics of a co-moving mass distribution in the background of a BST. This might throw light on some important local aspects of galactic dynamics.

\begin{center}
{\bf Acknowledgements}
\end{center}
It is a pleasure to thank Kanak Saha for very helpful conversations and email correspondence, and Geetanjali Sarkar for pointing out some useful references.

\end{document}